\documentclass[aps,prl, twocolumn]{revtex4}
\usepackage{amsmath}
\usepackage{amsfonts}
\usepackage{amssymb}
\usepackage{dcolumn}
\usepackage{bm}
\usepackage{graphicx}
\usepackage{epstopdf}
\usepackage{color}
\usepackage{ulem}

\def \Inpar {In$(2_{ac})$ }
 \def \InparE {In$(2_{ac})$}
\def \InperE {In$(2_{bc})$}
 \def \Inper {In$(2_{bc})$ }
\def\Ce {CeCoIn$_5 $ }

\begin{document}

\title{Field Evolution of Coexisting Superconducting and Magnetic Orders in \Ce  }

\author{ G. Koutroulakis$^{1}$, M. D. Stewart, Jr.$^{1}$, V. F. Mitrovi{\'c}$^{1}$, M. Horvati{\'c}$^{2}$, C. Berthier$^{2}$,
G. Lapertot$^{3}$, and J. Flouquet$^{3}$}
\address{$^{1}$Department of Physics, Brown University, Providence, RI 02912, U.S.A. \\
$^{2}$Laboratoire National des Champs Magn{\' e}tiques Intenses, CNRS,  B.P.
166, 38042 Grenoble
Cedex 9, France\\
$^{3}$CEA, INAC, SPSMS, CEA Grenoble, 38054 Grenoble Cedex 9,
France}

\date{ \today}

\begin{abstract}

We present  
nuclear magnetic resonance (NMR) measurements on the three distinct In sites of  \Ce with magnetic field   applied in the  [100] direction.
We identify the microscopic nature of the long range magnetic order (LRO) stabilized at low temperatures in  fields above 10.2 T while still in the superconducting  (SC) state.  We infer that the ordered moment is oriented along the  $\hat c$-axis and map its field evolution.
 The study of the  field dependence of the NMR shift for the different In sites indicates that the LRO likely coexists with a  modulated SC phase, possibly that  predicted by Fulde, Ferrell, Larkin, and Ovchinnikov. Furthermore, we discern a field region dominated by strong spin fluctuations where static LRO is absent  and propose a revised phase diagram.
  \vspace*{-0.5cm}
\end{abstract}

\maketitle

The detrimental effect of an applied magnetic field on a superconductor has nourished the intuitive presumption of a {\it de facto} competition between magnetic and superconducting (SC) orders. However, it is now well-established, both theoretically and experimentally, that not only can these orders coexist, but in some cases, they may even be essential for each other's stability \cite{Monthoux07}.  Manifestations of coexistence span a rather wide range of materials including several cuprates, ferropnictides, and heavy fermion systems of which \Ce is one of the most intriguing examples \cite{Flouquet06}. In the SC state   of this compound application of a magnetic field $(H_0)$ induces a long range magnetic order (LRO), restricted to a narrow low-temperature $(T)$ region of the phase diagram below $H_{c2}$ \cite{Young07, Kenzelmann08}. What is more, this particular region of the phase diagram was initially identified as the first realization of the long-sought Fulde, Ferrell, Larkin, and Ovchinnikov (FFLO) state, a superconducting state with a non-zero pair momentum and a spatially modulated order parameter \cite{FFLOdis,  MatsudaRev}. However, important questions regarding the true nature of the low-$T$ high-$H_0$ SC phase, the details of the magnetic order and its field dependence, and the potential driving mechanisms of their coexistence remain unanswered  \cite{CurroReview}. Thus, \Ce provides a strikingly rich  ground to study the complex interplay between exotic SC and magnetism. Experimentally, nuclear magnetic resonance (NMR), as a microscopic probe sensitive to both magnetic and SC degrees of freedom, provides a powerful tool for the investigation of these puzzles. 

In this letter,  detailed  low temperature
NMR  measurements on the three distinct In sites  in \Ce for $\mathbf{H_0} \, ||\, [100]$ are presented.
We establish that at $T \approx 70  \, {\rm mK}$ a phase with static magnetic   LRO  is stabilized for   fields above $\approx 10.2  \, {\rm T}$ in the SC state.  We deduce that the LRO is an incommensurate spin density wave (IC-SDW) with
moments  oriented along the    $\hat c$-axis, independent of the in-plane ${\mathbf H_0}$ orientation. Further, the detailed field evolution of the moment is mapped.
The study of the field dependence of the NMR shift  implies that this  IC-SDW
coexists with   a novel SC state, characterized by an enhanced spin susceptibility \cite{Mitrovic06}.   Finally,  we identify a new region in the \mbox{$H_0$-$T$} phase diagram, lying in between the low field   SC (lfSC) and the  IC-SDW states. This   could be an  FFLO phase
without magnetic LRO, in agreement with   recent theoretical predictions \cite{Yanase09}.

High quality single crystals  of CeCoIn$_5$,  grown by a flux method,   were placed
in NMR  radio frequency (RF) coils inside the mixing chamber of a dilution refrigerator so that  $\mathbf{H_0} \, ||\, [100]$. The RF coil was   used to determine the precise value of $H_0$ by performing $^{63}$Cu NMR on its copper nuclei. The spectra were obtained, at each given value of $H_0$, from the sum of spin-echo Fourier
transforms recorded at constant frequency intervals.
Extremely weak RF excitation power \cite{MitrovicConfP} was used to discern the NMR signal from In sites sensitive to magnetism.

\begin{figure}[t]
\begin{minipage}{0.98\hsize}
\centerline{\includegraphics[scale=0.53]{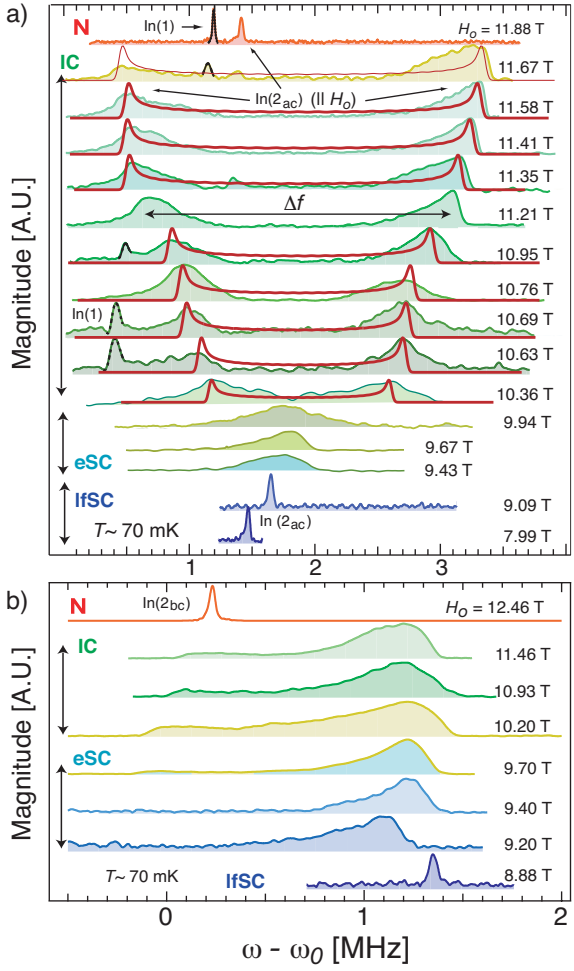}} 
\begin{minipage}{.98\hsize}
 \vspace*{-0.3cm}
\caption[]{\label{Fig1} (Color online)  NMR spectra of  \textbf{(a)}  In(1), $In(2_{ac})$ and \textbf{(b)}  $In(2_{bc})$ 
at \mbox{$T = 70$ mK} for various $\mathbf{H_0} \, \| \, \hat{a}$.  The frequency scale  is defined by subtracting $\omega_{0}$,  the zero NMR  shift frequency.
{\bf N} denotes  the normal phase, {\bf IC} the  LRO phase, {\bf eSC} the state with strong fluctuations, and {\bf lfSC} the Abrikosov SC state. 
 Solid lines in \textbf{(a)} are simulated spectra for the IC-SDW order described in the text.
}
 \vspace*{-0.6cm}
\end{minipage}
\end{minipage}
\end{figure}

 For  $\mathbf{H_0} \, ||\, [100]$, there are three inequivalent In sites. The  axially  symmetric In(1) is  located in the center of the tetragonal Ce planes,  while  \Inpar and  \Inper sites   correspond  to In atoms located on the lateral faces (parallel and perpendicular to the applied field, respectively) of the unit cell \cite{Curro01}.
In \mbox{Fig. \ref{Fig1}}  the   $H_0$  evolution  of the \Inpar and \Inper spectra at
\mbox{$T \sim$70 mK} is plotted.
  Lowering the field below \mbox{$\sim 11.7$ T} establishes magnetic  LRO.  The LRO is evident in the fact that the  \Inpar  line broadens into a spectrum with two extrema/peaks with finite signal weight in between them.  Such spectra are characteristic of   IC LRO  along one spatial dimension \cite{BlincIC}.
    Furthermore, at  \mbox{$H_0 = 11.67$ T}, both  the broad \Inpar   and the sharp normal state spectra are observed. This reflects the coexistence of normal and IC phases in the vicinity of the phase transition  confirming its first order character.
 From independent NMR measurements   \cite{Koutroulakis08} and   the tuning resonance of the   tank circuit, we establish that this transition to the IC LRO state coincides   precisely with  the transition from the normal to the SC state.
For  \mbox{$9.2 \, {\rm T} \lesssim H_0  \lesssim 10.2$ T},
the spectra of all In sites consist  of a single peak, {\it i.e.} no signature of the IC state is observed. However, these spectra
remain significantly broader than the ones for \mbox{$H_{0}\lesssim 9.2$ T}, where the linewidth of all sites can be adequately described by the spatial distribution of magnetic fields resulting from the  vortex lattice \cite{Koutroulakis08}.
%
%
 %
%

We now proceed to the analysis of the nature of the  IC LRO phase to elucidate  the field evolution  of its magnetic moment and its ordering wave vector ($\mathbf Q$).  The NMR spectrum reflects
$\mathcal{P}(H^{\|}_{int})$, the  probability distribution of the internal magnetic field  at the nuclear site projected along the $\mathbf{H}_0$
direction, denoted by  $\hat z$. In the LRO phase, the component of the internal hyperfine field parallel to $\mathbf{H}_0$, at an  In  site,
is given by
$H^{\|}_{int}  = \hat z \cdot \sum_{\left<i\right>} \mathbb{A}_i\cdot \boldsymbol{\mu}_i ,$
where  $\mathbb A_{i}$ is the symmetric $3 \times 3$ hyperfine coupling tensor with  the i$^{th}$ nearest-neighbor Ce  atom and $\boldsymbol{\mu}_i$ is
its magnetic moment \cite{ItinerElect}.  The exact form of the hyperfine tensor is derived    in \mbox{Ref. \cite{Curro06}}. Our data and the fact
that certain off-diagonal elements of  $\mathbb{A}$ ($ A_{  a   b}$, $  A_{  a   c}$, and $  A_{  b   c}$ for In(1), \InparE, and \InperE, respectively)    are non-zero allows us to place stringent constraints on the possible nature of the magnetic LRO.    Specifically, the form of $\mathbf Q$ and $\boldsymbol{ \mu}_i$ should be such that
the \Inpar spectrum (\mbox{Fig. \ref{Fig1}a})  broadens into a  double peak structure,
while no such broadening of the  In(1) \cite{Koutroulakis08} and  \Inper (\mbox{Fig. \ref{Fig1}b})   lines  is induced.  In principle, such spectra can be
effectively  described by    an appropriate  sinusoidal variation of  $H^{\|}_{int}$,  that is,  of the moment itself. For an IC-SDW state, this spatial
variation of the moment can be written as
$\boldsymbol{ \mu}_i (\mathbf r) = \boldsymbol{ \mu}_{0} \cos(\mathbf Q \cdot \mathbf r +\phi_0)$ where $\mathbf Q$ is the IC wave vector,  
$\mathbf r$ defines the lattice coordinates of the ${\rm i}^{th}$  Ce moment,  and   $\phi_0$ is an arbitrary phase.    The separation between  the
two   
extrema of the  \Inpar spectrum  is  $\Delta f = \gamma (H_{int}^{|| max} - H_{int}^{|| min}   )$, where $\gamma$ is the nuclear gyromagnetic ratio.   Thus,   $\Delta f$
depends on a product of hyperfine tensor components, the   moment  amplitude $\mu_{0}$, and     some  trigonometric function  dictated by $\mathbf Q$.

{\it A priori}, the  magnetic structure with the ordered moments
$\boldsymbol{\mu}_i\| \mathbf{H}_{0} \| \hat{a}$ and $ \mathbf{Q}
= \mathbf{Q_{\rm AF}}+\boldsymbol{\delta} $, where $\mathbf Q_{\rm
AF} = (0.5, 0.5, 0.5)$ and  incommensuration $\boldsymbol{\delta}=
(\delta, 0, 0)$, can satisfy the requirements  imposed by our
data, as was suggested in \cite{Young07}. In this case, $\Delta f$
for the   \Inpar peaks  is equal to  $4\gamma A_{{aa}}\mu_{0}\sin
(\delta \pi)$, with $A_{{aa}} \approx A^{LT}_{{\rm In}(2_{ac})}
/2\approx -0.6 \,{\rm T}/\mu_{B}$ \cite{Curro01, Curro06}.  
However,  reproducing the  observed  spectra  requires  either a
moment $\mu_{0}$ that is three times larger than that found by
neutrons   for  $ \mathbf{H}_{0} \| [1 \overline{1} 0]$
\cite{Kenzelmann08}  or an excessive value of $\delta \simeq
0.21$. It is thus unlikely that such a structure with
$\boldsymbol{\mu}_i || \hat a$  is stabilized in the IC  phase.

A better candidate for describing our data is  an IC-SDW with ordered moments   perpendicular to the plane,  {\it i.e.}
 $\boldsymbol{\mu}_i \| \hat c \perp\mathbf{H_{0}}$ as in   \cite{Kenzelmann08},  and $\boldsymbol{\delta}$ in  the plane.
  We cannot  uniquely determine the   direction of   $\boldsymbol{\delta}$, since the relevant $\mathcal{P}(H^{\|}_{int})$ is nearly  the same
  for $\boldsymbol{\delta}\| \hat{a}$, $\boldsymbol{\delta}\| \hat{b}$, or $\boldsymbol{\delta}\| [110]$.
However, the case of $\boldsymbol{\delta}\| \hat{b}\perp \mathbf{H}_0$ appears to be the most plausible one, since it would agree with both
the experimental  findings  for $\mathbf{H_0} \, ||\, [1 \overline{1} 0]$   \mbox{\cite{Kenzelmann08}}  and the theoretical prediction of
 a $\boldsymbol{\delta} \perp \mathbf{H}_0$  for coexisting IC-SDW and FFLO order parameters  \cite{Yanase09}.
In this case, $\Delta f$  is proportional to
 $4 \gamma A_{ac}  \mu_0$, where   $A_{{ac}}$ is  the off-diagonal hyperfine tensor component, whose value is not known from other independent  measurements. Assuming that at \mbox{11 T}  $\mu_{0}  = 0.15 \mu_{B}$  and
$| \boldsymbol\delta| = 0.085$ (the same magnitude as in  \cite{Kenzelmann08}) we find that a reasonable value of $A_{ac} \approx 0.38 \,{\rm T}/\mu_{B}$ is required to fully account for our data.
The calculated  distributions $\mathcal{P}(H^{\|}_{int})$,
convolved with the underlying vortex lattice lineshape \cite{Koutroulakis08}, are depicted
as the solid lines in \mbox{Fig. \ref{Fig1}a}.    Based on this result, the magnetic
LRO for $\mathbf{H}_0 || \hat a$ is most likely an IC-SDW   with the ordered moment
perpendicular to the   plane. Hence, in conjunction with the neutron  findings \cite{Kenzelmann08}, we infer that {\it the direction of the ordered moments is independent of the in-plane orientation of $H_0$}.

\begin{figure}[t]
\begin{minipage}{0.98\hsize}
\centerline{\includegraphics[scale=0.39]{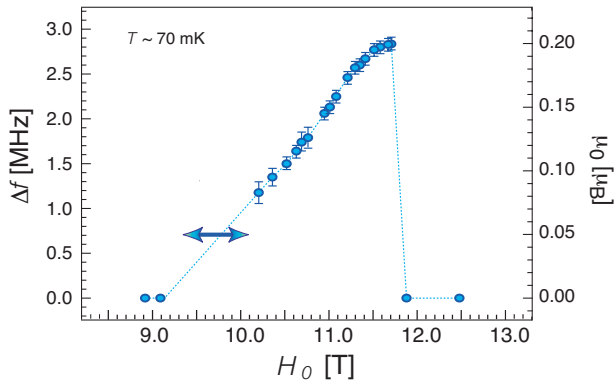}} 
\begin{minipage}{.98\hsize}
\vspace*{-0.3cm}
\caption[]{\label{Fig2} (Color online)
Magnetic moment as a function of $H_0$   deduced from $\Delta f$ at $T \approx 70 \, {\rm mK}$.
The double arrow denotes the region of $H_{0}$ within which LRO   is absent but spectra remain much broader than in the lfSC state.  }
\vspace*{-0.5cm}
\end{minipage}
\end{minipage}
\end{figure}

Having inferred that the magnetic moment is oriented perpendicular to the plane, we proceed to extract  the field evolution of its magnitude $\mu_0$ from $\Delta f$ of the \Inpar   depicted in   \mbox{Fig. \ref{Fig1}a}.
To do so, we fit the data to $\mathcal{P}(H^{\|}_{int})$
  with $\mu_0$ as a fitting parameter and assuming that   $|\boldsymbol \delta|=0.085$ is $H_{0}$ independent, as experimentally found  \cite{Kenzelmann08} and theoretically predicted   \cite{Miyake08, Yanase09}, and that the
   hyperfine component $A_{ac} \approx 0.38 \,{\rm T}/\mu_{B}$. 
   The deduced field evolution of $\mu_0$  is shown in  \mbox{Fig. \ref{Fig2}}.
The moment is zero outside the IC phase below $H_{0} \approx 10.2 \, {\rm T}$.
 It  
 increases by   a factor of $\sim 2.5$ as the field changes from
\mbox{$\approx 10.3$ T} to \mbox{11.67 T}, where it reaches its maximum value of \mbox{$\approx 0.2 \mu_{B}$}.

\begin{figure}[t]
\begin{minipage}{0.98\hsize}
\centerline{\includegraphics[scale=0.385]{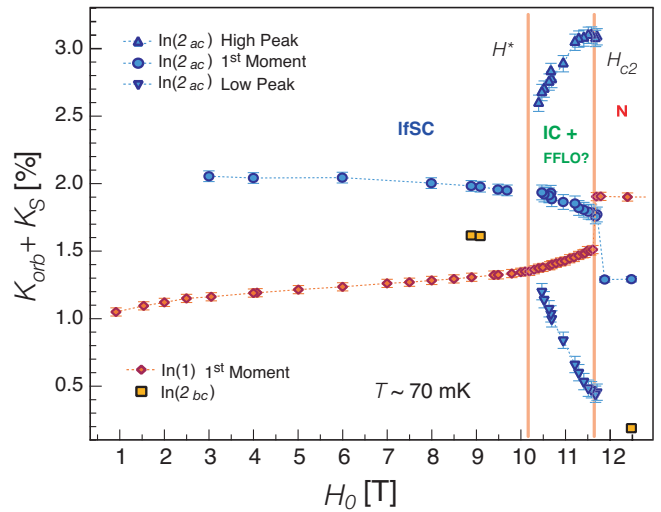}} 
\begin{minipage}{.98\hsize}
\vspace*{-0.3cm}
\caption[]{\label{Fig4}  (Color online)  NMR shift (spin, $K_s$, and constant orbital part, $K_{orb}$ \cite{Curro01}) of In(1), \Inpar and  \Inper at \mbox{$T \approx  70$ mK} as a function of $\mathbf{H_0} \|\hat{a}$. The solid lines indicate the values of $H^*$ and $H_{c2}$. 
}
\vspace*{-0.6cm}
\end{minipage}
\end{minipage}
\end{figure}

In order to  better understand the perplexing relationship between the coexisting SC and IC-SDW orders, we next discuss the field dependence of the NMR shift of \Inpar and compare it to that of In(1)  \cite{Koutroulakis08}.  This allows us to
distinguish between the contribution  from the low energy local
density of states (LDOS), characteristic of the SC state, which
affects both sites and that from the localized magnetic
moments,  affecting predominantly \InparE.
Our previous  analysis of the In(1) shift has revealed that a novel  SC
state, possibly an FFLO state,
is stabilized for   $H_0 = H^{*}\gtrsim  10$ T   via  a  \mbox{$2^{\rm nd}$ order} phase transition \cite{Koutroulakis08}.
We were, however,  unable to exclude  the existence of  a  magnetic order in this novel SC phase.
In  \mbox{Fig. \ref{Fig4}},   the field dependence of the shift,  determined by   diagonalizing  the full nuclear spin Hamiltonian, is plotted.    In the lfSC phase,
 the DOS  increases with $H_{0}$ due to excess Zeeman and Doppler-shifted nodal quasiparticles \cite{Koutroulakis08}, leading to an increasing spin susceptibility $(\chi_s)$.
The  In(1) and \Inpar   shifts ($K\propto A  \chi_{s}$)     exhibit reversed $H_0$ evolution. That is, they    scale with their respective hyperfine coupling constants, which are nearly equal but   have opposite  signs  \cite{Curro01}, with that for  \Inpar  being negative.
The same is also true for the relative shift change between the normal and lfSC state for both sites.

For  $H^{*} \le H_0 \le H_{c2}$,  where
 the IC-SDW LRO is established, we
    plot  the  \Inpar   shift   of  
    its first moment  and of the two extrema of its broadened lineshape.
   The difference of the shift of the two  extrema   
    strongly varies with the field reflecting the increase of $\Delta f$ with increasing $H_0$.
Further, the   \Inpar  shift associated with  its first moment    has  essentially the same field dependence as that reported for In(1),
once the difference in their  hyperfine couplings is taken into account.
{\it Thus,  independently of  the local magnetism, both In sites
sense an additional common local field, due to a spin polarization
along ${\mathbf H}_0$.}

There are two likely origins for this
 local  field   for $H_{0}  \ge H^{*}$.   One  is a canting of the transverse staggered
magnetization of the localized moments along
  ${\mathbf H}_0$. 
To test this hypothesis one  needs a quantitative description of  canting in an IC phase, which is  missing. Thus,  we  calculate the effect of several crude models of canting   on the NMR observables.  
We find that our data are effectively  reproduced by
considering an appropriate fixed value of $\mu_{\hat z}$ $(\sim \mu_0 \sin (5^{\circ}))$ on all lattice sites, 
regardless of $\left| \boldsymbol{\mu}_i(\mathbf r )\right|$. Nevertheless, it is not   clear  how this type of canted moment 
could be induced    in an IC-SDW phase.

  Alternatively, the additional  common field  could originate from  the enhanced LDOS  of the spin polarized quasiparticles in the   nodal planes  of the   FFLO state \cite{Mitrovic06}.  In that case, since in an FFLO state LDOS varies on a length scale that significantly exceeds the spacing between two In sites, its contribution to the shift of the two distinct In sites should be equivalent as observed.
This spatial modulation of the  LDOS   can also contribute to asymmetric  spectral broadening, which  for different In sites should scale as their respective hyperfine couplings.
For  In(1), this LDOS modulation gives rise to a tail on the high frequency side of the line  \cite{MitrovicConfP, Koutroulakis08}, while for \Inpar  the tail should be on the low frequency side, due to the sign difference of their   hyperfine coupling.
This is indeed observed for
\Inpar spectra, where only the low frequency side  is  essentially  broadened beyond   IC-SDW lineshape as shown in \mbox{Fig. \ref{Fig1}a}.

Next, we  consider the intriguing  low field limit of the IC LRO phase.
The low-$T$  shift data provide    evidence of a continuous  phase transition at $\approx 10.2$ T,  as shown in \mbox{Fig. \ref{Fig4}}.
However, significant broadening of   \Inpar and \Inper spectra, as compared to the ones in the lfSC phase, onsets for $H_{0}  \gtrsim 9.2 \, {\rm T}$, as evident in  \mbox{Fig. \ref{Fig1}}.  Strikingly, 9.2 T is precisely the field at which the amplitude of the IC-SDW should vanish if one extrapolates from its
field dependence in \mbox{Fig. \ref{Fig2}}.  For $9.2\, {\rm T}\, \lesssim H_{0}\lesssim10.2 \, {\rm T}$,   the  \Inpar signal  is   weak
 due to the rapid loss of   spin coherence, caused by strong    field fluctuations. Thus, although the static LRO is absent,  strong antiferromagnetic (AF) fluctuations are still present. These fluctuations
 can be responsible  for the apparent collapse, evident in the disappearance of the double-horned    broad \Inpar line, of the LRO in this field region.
  Besides,   the observed additional static  broadening, as compared to the lfSC state,    in the presence of these fluctuations,  can only emerge from
  highly enhanced $\chi_{s}$ around defects \cite{Berthier78}, such as  nodal planes in an FFLO state.  Thus,  this field range possibly corresponds to an FFLO state in the presence
of strong AF fluctuations, as   predicted in \cite{Yanase09}.
This is all   more likely since the width of
the \mbox{low-$T$} spectra of all three In sites in 
 the same field  range  is comparable to that observed in
the higher $T$ $(T> T^{*})$, high $H_{0}$ $(H_{0} > H^{*})$ SC phase outside the limits of the LRO state   \cite{MitrovicConfP,Young07}.

\begin{figure}[t]
\begin{minipage}{0.98\hsize}
\centerline{\includegraphics[scale=0.40]{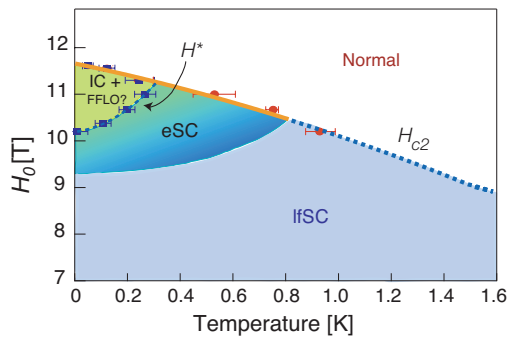}} 
\begin{minipage}{.98\hsize}
\vspace*{-0.3cm}
\caption[]{\label{Fig3} (Color online)  A sketch of an alternative   phase diagram of CeCoIn$_{5}$ (data points adopted from \mbox{Ref. \cite{MitrovicConfP}}).
The solid and dashed lines indicate
   $1^{{\rm st}}$ and  $2^{{\rm nd}}$ order phase transitions, respectively.
}
\vspace*{-0.6cm}
\end{minipage}
\end{minipage}
\end{figure}

The 2$^{\rm nd}$ order phase transition previously  identified by the  In(1) shift  can   be, then, the transition to the IC LRO  state. As recently shown,    the IC magnetism can arise in the FFLO state  in a $d$-wave SC as a consequence of the formation of   Andreev bound states near the zeros of the FFLO order parameter \cite{Yanase09}. It is a large LDOS in the bound states that triggers the formation of the IC LRO \cite{Yanase09cp}. At finite $T$  the LRO   phase is stabilized only when a sufficient number of the nodal planes  containing the bound states  is induced by $H_{0}$ \cite{Yanase09, NoteAFpt}. Since  these same  states  contribute to the  shift, it is
likely that the transition   identified by it  is indeed   the one to the IC LRO  phase.   Alternatively, if the visible increase in the In(1) shift as compared to that in  the lfSC phase were due to canting,
an evident increase   should onset  only  after LRO is established.

Based on these observations and consistent with the discussion in the previous paragraphs,   we  postulate  that
a phase with strong AF  fluctuations  (referred to as  `exotic' SC (eSC)) develops for $9.2 \, {\rm T}\, \lesssim H_{0}\lesssim10.2  \, {\rm T}$ at low-$T$, while
the $2^{\rm nd}$ order phase transition previously reported \cite{Bianchi03,Koutroulakis08, MatsudaRev} at \mbox{$H^{*} \approx 10.2$ T} marks the transition to the state with well established static IC-SDW order.
Such a two step phase transition from lfSC to FFLO  (assuming it exists in eSC) and then to the IC state was theoretically predicted when spin fluctuations  are considered in \mbox{Ref. \cite{Yanase09}}.  In light of our NMR results, we propose the revised phase diagram   sketched  in \mbox{Fig. \ref{Fig3}}.

In conclusion, our comprehensive low-$T$ NMR data provide a  clear  picture   of the
field evolution of the IC magnetism, and the magnitude of the magnetic moment,  confined within the high-$H_0$  low-$T$ phase of  CeCoIn$_{5}$.
In conjunction with neutron scattering results, we deduce that the direction of the ordered moment    does not depend on the in-plane applied field  orientation.
Our analysis of  the  field dependence of the NMR shift on different In sites indicates that the IC-SDW order likely coexists with an FFLO state. Finally, we identify a novel phase,  in the  field regime in
between the lfSC   and IC-SDW states,  which is   likely a true FFLO phase in the presence of strong AF fluctuations.

Many thanks to S. Kr{\"a}mer for helping with the experiment. This research is
supported by the funds from NSF (DMR-0710551), ANR grant 06-BLAN-0111, and the
GHMFL, under European Community contract RITA-CT-2003-505474. V. F. M.
acknowledges support by the A. P. Sloan Foundation.

\vspace{-0.7cm}
\bibliographystyle{unsrt}

\begin{thebibliography}{10}
\bibitem{ItinerElect}
We assume that  itinerant electron contribution is  negligible compared to that from the static localized moments.
\end{thebibliography}

\vspace{0.0cm}
\end{document}